\documentclass[a4paper,12pt]{article}
\usepackage[T1]{fontenc}
\usepackage[utf8]{inputenc}
\usepackage{amsmath}
\usepackage{amsfonts}
\usepackage{caption} 
\usepackage{graphicx}
\usepackage{amssymb}
\usepackage{pdflscape}
\usepackage{array}
\usepackage{hyperref}
\usepackage{authblk}
\usepackage{mathtools,float}
\DeclareMathOperator{\Tr}{Tr}

\title{\bf{Supermultiplet of $\beta-$ deformations from twistors}}

\author[*]{Segundo P. Mili\'an}
\affil[*]{Instituto de Física Teórica, Universidade Estadual Paulista\\R. Dr. Bento Teobaldo Ferraz 271, Bloco II- Barra Funda \\
CEP:01140-070-São Paulo, Brasil}

\date{}
\begin{document}

\maketitle

\begin{abstract}
We consider the supermultiplet of linearized beta-deformation of $\mathcal{N}=4$ Super Yang-Mills(SYM). It was previously studied on the 
gravitational side.
We study the supermultiplet of beta-deformations on the field theory side and we compare two finite-dimensional representations of $psl(4|4,\bf{R})$ algebra.
We show that they are related by an intertwining operator. We develop a twistor-based approach which could be useful for studying 
other finite-dimensional and nonunitary representations in AdS/CFT correspondence.

\end{abstract}
\tableofcontents
\section{Introduction}

Much  progress in order to understand the AdS/CFT correspondence has been achieved since it was formulated \cite{Maldacena:1997re, Aharony:1999ti}. 
An important tool is the study of small perurbations. 
In the CFT side they correspond to deformations of the SYM action.

Small perturbations in the supergravity(SUGRA) side  correspond to small fluctuations of the classical SUGRA fields around their "vacuum" values in
 the maximally symmetric background solution, $AdS_5\times S^5$. There was extensive study of small fluctuations \cite{Kim:1985ez}. However, even at the linearized level, the research was mainly focused 
 on unitary representations \cite{Gunaydin:1984fk}. We feel that non-unitary representations are also important and have not been sufficiently studied in this context.
  In particular, there are finite-dimensional representations
\cite{Mikhailov:2011af}. To the best of our knowledge, the classification of finite-dimensional representations is missing.
   
The beta-deformation is probably the simplest example of    a finite-dimensional representation, it 
is very well studied on the SUGRA side \cite{Lunin:2005jy}, \cite{Frolov:2005dj}, \cite{Grassi:2006tj}, \cite{Bedoya:2010qz}. 
On the field theory side, it is   a particular case 
of the  deformation  studied by Leigh and Strassler \cite{Leigh:1995ep}. 
It preserves $\mathcal{N}=1$ SUSY and depends on two complex parameters. 

However, the full supermultiplet has never been studied on the field theory side.
And we fill this gap with the present paper. 
Moreover, we develop a new method, which is hopefully useful also
for other non-unitary representations.
Namely, we consider the deformation of the SYM action, the one considered in \cite{Leigh:1995ep}. It is 
evaluated on the tensor product of singleton representations. We will use the oscillator  representation of the singleton representation
of $psl(4|4, \bf{R})$ 
\cite{Bars:1982ep}; for a review of the superconformal algebra see \cite{Beisert:2010kp}.

The $\mathcal{N}=4$ SYM admits a formulation in twistor space \cite{Witten:2003nn}, see also \cite{Boels:2006ir}.
A vector in the singleton representation can be understood as a wave function of the free
field on ${\bf{R}}^{2,2}$ supported on the $\alpha$-plane. 
This method allows us to reduce the problem to finding the
fininite-dimensional invariant subspace in the tensor
product of three singleton representations. Basically, in order to classify deformations of the SYM action, it is enough to 
evaluate them on tensor products of the spaces of solutions of free equations, which are dual to singletons.

\vspace{.3cm}
{\bf{Plan of the paper}}
\;We make a review of $\mathcal{N}=4$ SYM theory and $psl(4|4,\bf{R})$ superalgebra. Namely, we focus on the oscillator representation of 
$psl(4|4,\bf{R})$, which we call
oscillator picture. We also review the relation between the oscillator picture and supertwistors. We discuss how to write single trace operators on both pictures in Section ~\ref{II}. Then in Section
~\ref{III} we present the structure of the supermultiplet of $\beta-$deformation on the field theory side. Using the oscillator 
picture  we show the conformal invariance of $\mathcal{O}[B]$ in Appendix ~\ref{a}.

\section{Marginal Deformations in $\mathcal{N}=4$ SYM theory and $psl(4|4,{\bf{R}})$}
\label{II}
 ${\mathcal{N}=4}$ SYM is a field theory  with the maximal amount of supersymmetry in four dimensions and is unique \cite{Grimm:1977xp}, for an extended
 review see Chapter 3 of \cite{D'Hoker:2002aw}. The action of this field theory is 
 \begin{eqnarray}
\label{N=4SYM}
 S_4
 &=&-\int d^4x\big(\frac{1}{4}F^{(a)}_{\mu\nu}F^{\mu\nu(a)}+\frac{1}{4}D_{\mu}\phi^{AB(a)}D^{\mu}\phi_{AB}^{(a)}+i\bar\psi^{(a)}_A\bar\sigma^{\mu}D_{\mu}\psi^{(a)A} +\frac{\sqrt{2}}{2}g_{YM}f_{abc}\psi^{\alpha A(a)}\phi_{AB}^{(b)}\psi_{\alpha}^{B(c)} \nonumber \\
 &+&\frac{g^2_{YM}}{16}f_{abc}f_{ade}\phi^{AB(b)}\phi^{CD(c)}\phi_{AB}^{(d)}\phi_{CD}^{(e)} -\frac{\sqrt{2}}{2}g_{YM}f_{abc}\bar\psi^{(a)}_{\dot\beta A}\phi^{AB(b)}\bar\psi^{\dot\beta (c)}_{B} \big), 
\end{eqnarray}
where capital letters run from $1$ to $4$.

$\mathcal{N}=4$ SYM admits a formulation in terms of $\mathcal{N}=1$ language, one of the four fermions is combined with the gauge field 
to form the vector superfield. The remaining fermions are combined with three complex scalar, to form three chiral superfield. 
The superpotential $\mathcal{W}$ is given by  

\begin{equation}
\label{W}
 \mathcal{W}=\frac{g_{YM}\sqrt{2}}{3!}\epsilon_{ijk}f_{abc}\Phi^i_a\Phi^j_b\Phi^k_c,
\end{equation}
where $i,j,k$ indices running from $1$ to $3$, $\Phi^i$ denote the three chiral superfields and $a,b,c$ denote gauge group SU(N) indices.

In this work we are interested in study deformations of this field theory. To begin with, let us start with the deformation studied by  Leigh and Strassler 
~\cite{Leigh:1995ep}. The superpotential ~\eqref{W} is deformed to
\begin{eqnarray}
 \mathcal{W}\mapsto \mathcal{W}+\frac{1}{6}h_{ijk}d_{abc}\Phi^i_a\Phi^j_b\Phi^k_c,
\end{eqnarray}
where $d_{abc}=\Tr(T_a\{T_b,T_c\})$. The  action at linear order in $h$ is 
\begin{eqnarray}
 S=S^{\text{free}}_4-\frac{1}{4}\int d^4x B^{[MN]}_{(IJ)}d_{abc}\psi^{(a)I}\psi^{(b)J}\phi_{MN}^{(c)}+\text{h.c.} \phantom{-},
\end{eqnarray}
the deformation is conformal invariant at classical level for all values of $h_{ijk}$.  $S^{\text{free}}_4$ denotes the action of $\mathcal{N}=4$ SYM in free field theory. Here the coupling $B^{[MN]}_{(IJ)}$ obeys 
the tracelessness condition ~\cite{Bedoya:2010qz} 
\begin{eqnarray}
 B^{[MN]}_{(IN)}&=&0, 
\end{eqnarray}
where $B^{[mn]}_{(ij)}=h_{ijl}\epsilon^{lmn}$, with latin indices running from $1$ to $3$. We will see that this condition is crucial for matching 
with the field theory side.%
\subsection{The superalgebra $psl(4|4, \bf{R})$ }
The symmetry group of the action~\eqref{N=4SYM} is  $PSU(2,2|4)$. Local operators of $\mathcal{N}=4$ SYM are arranged in multiplets of the algebra
$psl(4|4, \bf{R})$ \cite{Dobrev:1985qv}.
The generators of  the algebra $psl(4|4,\bf{R})$ are the $sl(2)\times sl(2)$ rotations $l_{\beta}^{\;\alpha},\; \dot l_{\dot\beta}^{\;\dot\alpha } $, the $sl(4)$ rotations, the translation $p_{\alpha\dot\alpha}$, the 
conformal generator $k^{\dot\alpha\alpha}$, %
the supersymmetry generators $q_{A\alpha}$, $\bar q^{A}_{\dot\alpha}$, the superconformal generators $s^{A \alpha}$, $\bar s^{\dot\alpha}_{A}$ as well as 
dilatation generator $D$. 

Using  $\mathcal{N}=1$ language to write down the action $S_4$, it
breaks the original  ${\mathcal{R}}$ symmetry into 
$SU(3)\times U(1)$ .

\subsubsection{Oscillator representation of $psl(4|4, \bf{R})$ superalgebra}%
\label{oscrepres}%

The oscillator method was developed in order to construct unitary irreducible representations of non-compact groups \cite{Bars:1982ep} 
in terms of its maximal compact subgroup. 
Namely, the generators of the $gl(4|4, \bf{R})$ superalgebra can be represented in terms of two sets of bosonic oscillators $(a^{\alpha}, \;a^\dagger_{\alpha})$, $(b^{\dot\alpha}, b^\dagger_{\dot\alpha})$ with $\alpha, \;\dot \alpha=1,2$ and one set of fermionic oscillator 
$(c_{A}, \;c^{\dagger A})$ with $A=1,2,3,4$. The non-vanishing relations of commutation are%
\begin{eqnarray}
[a^{\alpha}, a^\dagger_{\beta}]=\delta^{\alpha}_{\phantom{-}\beta}, \phantom{--}  [b^{\dot\alpha},b^\dagger_{\dot\beta}]&=&\delta^{\dot\alpha}_{\phantom{-}\dot\beta} \phantom{--} \{c_{A},c^{\dagger B}\}=\delta_A^{\phantom{-}B}.
\end{eqnarray}

The  $gl(4|4, \bf{R})$ generators can be written as products of two oscillators, the supercharges, translations and conformal transformations are %
\begin{eqnarray}
\label{off-diagonal-generators}
q_{A\alpha}&\;=\;&c_{A}a^\dagger_{\alpha}, \;\; \phantom{--}  s^{A\alpha}\;=\;c^{A\dagger}a^{\alpha}, \nonumber \\
\bar q^{A}_{\dot\alpha}&=&c^{A\dagger}b^{\dagger}_{\dot\alpha}, \;\; \phantom{--}\bar{s}^{\dot\alpha}_{ A}\;=\;b^{\dot\alpha}c_{A}, \nonumber \\ 
p_{\alpha\dot\alpha}&=&a^\dagger_{\alpha}b^\dagger_{\dot\alpha}, \;\; \phantom{--} k^{\dot\alpha\alpha}\;=\;b^{\dot\alpha}a^{\alpha}.
\end{eqnarray}
 The dilatation generator reads as follows
 \begin{align}
  D\;=\; 1+\frac{1}{2}a^\dagger_{\gamma}a^{\gamma}+\frac{1}{2}b^\dagger_{\dot\gamma}b^{\dot\gamma},
 \end{align}
and the $sl(2)\times sl(2)$ and $sl(4)$ rotation  generators are %
\begin{eqnarray}
\label{compact-generators}
 l_{\beta}^{\;\;\alpha}&=&a^\dagger_{\beta}a^{\alpha}-\frac{1}{2}\delta_{\beta}^{\;\alpha}a^{\dagger}_{\gamma}a^{\gamma}, \nonumber \\
 \bar l_{\dot\beta}^{\;\;\dot\alpha}&=&b^\dagger_{\dot\beta}b^{\dot \alpha}-\frac{1}{2}\delta^{\;\dot\alpha}_{\dot\beta}b^\dagger_{\dot\gamma}b^{\dot\gamma}, \nonumber \\
 r^{A}_{\;\; B}&=&c^{\dagger A}c_B-\frac{1}{4}\delta^{A}_{\;B}c^{\dagger D}c_{D}.
\end{eqnarray}
Also there are two $gl(1)$ generators%
\begin{eqnarray}
\label{gl-1 generators}
 C&=&\;1-\frac{1}{2}a^\dagger_{\gamma}a^{\gamma}+\frac{1}{2}b^\dagger_{\dot\gamma}b^{\dot\gamma} -\frac{1}{2}c^{\dagger D}c_D \nonumber \\
 B&=&\;-1+\frac{1}{2}a^\dagger_{\gamma}a^{\gamma}-\frac{1}{2}b^\dagger_{\dot\gamma}b^{\dot\gamma} ,
\end{eqnarray}
which are the central charge and outer derivation, respectively. 

All fields in $\mathcal{N}=4$ SYM are uncharged with respect to the central charge $C$, therefore can be dropped. This procedure leads to  $sl(4|4, \bf{R})$. The 
generator $B$ does not appear in commutators in $sl(4|4,\bf{R})$ and can be projected out, giving  the algebra $psl(4|4, \bf{R})$.

\subsection{$\mathcal{N}=4$ SYM fields in terms of oscillators and supertwistors}
Here we give a brief description of how we can write down solutions of free SYM theory in both oscillator representation and supertwistors. Also we present an equivalence between oscillator and supertwistor pictures. We finish this section writing composite operators in both pictures. 

To write down  the free field components of $\mathcal{N}=4$ SYM in the oscillator picture. 
Let be $|0\rangle$ an invariant  non-physical vacuum  under $psl(4|4,\bf{R})$, it is annhilated by $c_A$, $b^{\dot\alpha}$ and $a^{\alpha}$ \cite{Gunaydin:1984fk}. 

Let us define scalar fields as
\begin{eqnarray}
\label{scalar}
 \phi^{AB}\;=\;c^{\dagger A}c^{\dagger B}|0\rangle,
\end{eqnarray}
the another fields without derivatives can be obtained by applying the SUSY's generators $q_{A\alpha}$ and $\bar q^{A\dot\alpha}$ on ~\eqref{scalar}. They  read as follows
\begin{eqnarray}
\label{fermions}
 \psi^D_{\alpha}\sim a^\dagger_{\alpha}c^{\dagger D}|0\rangle \phantom{--} , \phantom{--}\bar\psi_{M\dot\alpha}\sim \epsilon_{MNBD}b^{\dagger}_{\dot\alpha}c^{\dagger N}c^{\dagger B}c^{\dagger D}|0\rangle,
\end{eqnarray}

and the self-dual and antiself-dual of the field strength are given by
\begin{eqnarray}
\label{sd-asd}
 f_{\alpha\beta}&\sim&a^{\dagger}_{\alpha}a^{\dagger}_{\beta}|0\rangle \phantom{--} \text{and} \phantom{--} \bar f_{\dot\alpha\dot\beta}\sim \epsilon_{ABDE}b^{\dagger}_{\dot\alpha}b^{\dagger}_{\dot\beta}c^{A\dagger}c^{B\dagger}c^{D\dagger}c^{E\dagger}|0\rangle,
\end{eqnarray}
where the tildes means up to a proportionality constant.%
\subsubsection{From oscillators representation to  supertwistors}%
The set of oscillators introduced above can be written in terms of supertwistors variables. %
Supertwistors $\mathcal{Z}$ parametrize the space $\bf{RP}^{3|4}$ \cite{Witten:2003nn}
\begin{align}
\mathcal{Z}\;=\; 
\left(
\begin{array}{c}
\lambda_{\alpha}\\
\mu^{\dot\alpha}\\
\psi^A
\end{array}
\right),
\end{align}
where $\lambda_{\alpha}$ and $\mu^{\dot\alpha}$ are two-components bosons  and $\Psi^A$ is a four-components fermions. A twistor $(\lambda_{\alpha},\mu^{\dot\alpha})$ 
defines a two-dimensional isotropic subspaces of ${\bf{R}}^{2,2}$ 
\begin{align}
\mu^{\dot\alpha}+\lambda_{\alpha}x^{\dot\alpha\alpha}\;=\; 0, 
\end{align}
which are called $\alpha-$planes. In the supersymmetric case,  $\psi^A$ define also a plane in $\theta$ space 
\begin{align}
\psi^A +\theta^{A\alpha}\lambda_{\alpha}\;=\;0,
\end{align}
where $x^{\dot\alpha\alpha}$ and $\theta^{A\alpha}$ are coordinates in superMinkowski spacetime
$\mathbb{\bf M}^{4|8}$. 

The oscillators  are related to the  the variables $\lambda, \, \mu$ and $\psi$, say  the twistor picture, by %
\begin{align}
a^{\dagger}_{\rho}\;\rightarrow\; \lambda_{\rho} , \;\;b^{\dot\alpha }\rightarrow\mu^{\dot\alpha}, \; \;\text{and}\;\; c^{\dagger A}\;\rightarrow\;\psi^A,
\end{align}
and so on. The generators of  $psl(4|4, \bf{R})$ can be written  in 
terms of supertwistor variables  leading to first-order differential operators ~\cite{Witten:2003nn}.

On-shell $\mathcal{N}=4 $ SYM fields can be described by a  scalar superfield $\Phi(\lambda,\mu,\eta)$ \cite{Berkovits:2009by}:
\begin{align}
 \Phi(\mathcal{Z})\;=\;\tilde f+\eta_A\tilde \psi^A+\frac{1}{2!}\eta_A\eta_B\tilde \phi^{AB}+\frac{1}{3!}\eta_{A}\eta_{B}\eta_{C}\epsilon^{ABCD}\tilde {\bar\psi}_D+\eta_{1}\eta_{2}\eta_{3}\eta_{4}\tilde {\bar f}.
\end{align}
Notice that $\tilde f$ and $\tilde{ \bar f}$ are independent of $\eta$ and that all the fields are in the twistor picture. The scalar $\Phi(\mathcal{Z})$  in twistor languague ~\cite{Adamo:2011cb,Cachazo:2004kj} is written as 
\begin{align}
\label{onshell-twistor}
\Phi(\mathcal{Z})\;=\;2\pi i\int_{\mathbb{C}}\frac{ds}{s}e^{s(\mu^{\dot\alpha}\bar p_{\dot\alpha}+\eta_A\xi^A)}\bar\delta^2(s\lambda_{\alpha}-p_{\alpha}),
\end{align}%
see Appendix C of ~\cite{Adamo:2011cb} for a derivation of~\eqref{onshell-twistor}
In order to recover the spacetime dependence, we need to Penrose transform $\tilde f,\; \tilde \psi^A,\; \tilde \phi^{AB}, \; \tilde{\bar \psi}$ and $\tilde{\bar f}$.
For a review  see appendix A of ~\cite{Witten:2003nn}.%

The scalar field~\eqref{scalar} is given by %
\begin{align}
 \phi^{AB}(x)\;=\;\int \frac{\lambda^{\alpha}d\lambda_{\alpha}}{2\pi i}\left.\frac{\partial}{\partial \eta_A}\frac{\partial}{\partial \eta_B}\Phi(\mathcal Z)\right\vert_{\eta=0},
\end{align}
roughly speaking, we can state that the non-physical vacuum $|0\rangle$ can be related to~\eqref{onshell-twistor} up to Penrose transform. The another  free fields~\eqref{fermions} and~\eqref{sd-asd}  are given as follows 

\vspace{.3cm}
%
\begin{small}
\begin{center} 
\begin{tabular}{| m{4em}| m{13em}|  m{21em}|}
\hline \vspace{0.3cm}
 Fields & Oscillators &  Stwistors\\
\hline \vspace{0.9cm}
$\Psi^{D}_{\alpha}(x)$ &$\mathcal{R}a^\dagger_{\alpha} c^{\dagger D}|0\rangle$ & $\int \frac{\lambda^{\alpha'}d\lambda_{\alpha'}}{2\pi i}\lambda_{\alpha}\left.\frac{\partial}{\partial \eta_D}\Phi(\mathcal Z) \right\vert_{\eta=0}$ \\
\hline \vspace{0.9cm}
$\bar\psi_{M \dot\alpha}(x)$ &$\epsilon_{MNBD}\mathcal{R}b^{\dagger}_{\dot\alpha}c^{\dagger N}c^{\dagger B}c^{\dagger D}|0\rangle$ & $\epsilon_{MNBD}\int \frac{\lambda^{\alpha'}d\lambda_{\alpha'}}{2\pi i}\frac{\partial}{\partial \mu^{\dot\alpha}}\left.\frac{\partial}{\partial \eta_N}\frac{\partial}{\partial \eta_B}\frac{\partial}{\partial \eta_D}\Phi(\mathcal{Z})\right\vert_{\eta=0}$ \\
 \hline \vspace{0.9cm}
$f_{\alpha\beta}(x)$ & $\mathcal{R}a^{\dagger}_{\alpha}a^{\dagger}_{\beta}|0\rangle$ & $\int \frac{\lambda^{\alpha'}d\lambda_{\alpha'}}{2\pi i}\lambda_{\alpha}\lambda_{\beta}\left.\Phi({\mathcal{Z}})\right\vert_{\eta=0}$ \\
\hline \vspace{0.9cm}
$\bar f_{\dot\alpha\dot\beta}(x)$& $\epsilon_{ABDE}\mathcal{R}b^{\dagger}_{\dot\alpha}b^{\dagger}_{\dot\beta}c^{A\dagger}c^{B\dagger}c^{D\dagger}c^{E\dagger}|0\rangle$& $\epsilon_{ABDE}\int \frac{\lambda^{\alpha'}d\lambda_{\alpha'}}{2\pi i}\frac{\partial}{\partial \mu^{\dot\alpha}}\frac{\partial}{\partial \mu^{\dot\beta}}\left.\frac{\partial}{\partial \eta_A}\frac{\partial}{\partial \eta_B}\frac{\partial}{\partial \eta_D}\frac{\partial}{\partial \eta_E}\Phi(\mathcal{Z})\right\vert_{\eta=0}$ \\
\hline
\end{tabular}
\end{center}
\end{small}

\vspace{0.3cm}
where $\mathcal{R}=e^{ip\cdot x}$ is the translation operator.

\subsubsection{Single trace operators}%

However in gauge theories with gauge group $SU(N)$, e.g. $\mathcal{N}=4$ SYM.  We need the product of these fields leading to composite operators. In this work we are interested in
single trace composite operators.

In this section, we explain how to write down single trace operators in terms of oscillators. We set up the following notation.
\begin{enumerate}
   \item If $\psi_1$ and $\psi_2$ are fermions then $\psi_1\bullet \psi_2 = {1\over 2} (\psi_1\otimes \psi_2 - \psi_2\otimes \psi_1)$, \footnote{$\bullet$ means the super-symmetrized tensor product.}
   \item if $\psi_1$ and $\phi_2$ are bosons then $\phi_1\bullet \phi_2 = {1\over 2} (\phi_1\otimes \phi_2 + \phi_2\otimes \phi_2)$, and
   \item if $\phi$ is boson and $\psi$ is fermion then $\phi\bullet\psi = {1\over 2} (\phi\otimes\psi + \psi \otimes\phi)$.
\end{enumerate}
At linear order in $h $, say $\mathcal{O}[B]$. The multiplet 
in the oscillator picture is %
\begin{eqnarray}
\label{B-multiplet}
\mathcal{O}[B]&=& \int d^4x \prod_{j=1}^3 \mathcal{R}^{(j)} B^{[MN]}_{(IJ)} \Psi^{(IJ)}_{[MN]} ,
\end{eqnarray}%
where $\Psi^{(IJ)}_{[MN]}$ is given by %
\begin{equation}
\label{Psi}
\Psi^{(IJ)}_{[MN]} = \;a^{\dagger}_{[1} c^{\dagger (I}|0\rangle \bullet a^{\dagger}_{2]} c^{\dagger J)} |0\rangle \bullet \epsilon_{MNKL} c^{\dagger K} c^{\dagger L} |0\rangle. 
\end{equation}%
 The index $j$ denotes  in which singleton representation acts
the $\mathcal{R}^{(j)}$ operator.
~\eqref{B-multiplet} is  invariant under conformal transformation. In  Appendix ~\ref{a} we show this fact using the oscillator representation.%
\newpage
In the {\bf{twistor picture}}~\eqref{B-multiplet}  is given by 
\begin{eqnarray}
\label{B-yukawa-twistor}
 \mathcal{O}[B]\;=\; \int d^4x\; \prod^3_{i=1} \int \frac{\lambda^{\alpha'}_{(i)}d\lambda_{\alpha'(i)}}{2\pi i}B^{[MN]}_{(IJ)}\Psi^{(IJ)}_{[MN]},
\end{eqnarray}
where $\Psi^{(IJ)}_{[MN]}$ reads as
\begin{align}
 \label{Psi-twistor}
 \Psi^{(IJ)}_{[MN]} \;=\;\epsilon_{MNAB}\epsilon^{\alpha\beta}\lambda_{\beta(1)}\left.\frac{\partial}{\partial \eta_{I(1)}}\Phi^{(1)}(\mathcal Z)\right\vert_{\eta=0}\;\lambda_{\alpha(2)}\left.\frac{\partial}{\partial \eta_{J(2)}}\Phi^{(2)}(\mathcal Z)\right\vert_{\eta=0}\;\left.\frac{\partial}{\partial \eta_{A(3)}}\frac{\partial}{\partial \eta_{B(3)}}\Phi^{(3)}(\mathcal Z)\right\vert_{\eta=0}.
\end{align}
We see that~\eqref{B-yukawa-twistor} is finite on-shell. We can evaluate it on the product of three off-shell
fields and the integral is convergent \footnote{We thank Prof. Andrei  Mikhailov for making clear this point.}. This means that~\eqref{B-yukawa-twistor}\;
defines an element of the dual space to the tensor product
of three singleton representations.%
\section{The structure of the supermultiplet}
\label{III}
In this section we describe the full supermultiplet of $\beta-$deformation in the field theory side. Instead of working in both oscillator picture and/or twistor picture, we just work out in the oscillator picture.
\subsection{Descendants multiplets from $\mathcal{O}[B]$ }
It is known that  $(c^{\dagger I}c_{M}\wedge c^{\dagger J}c_{N})_0$, which is $(g\wedge g)_0$ corresponds to $\Psi_{[MN]}^{(IJ)}$ ~\cite{Bedoya:2010qz}. 

Below we list in a table the set of deformations arising from the multiplet $\mathcal{O}[B]$.
%
\begin{table}[htb]
\centering
\caption{Descendants from $\mathcal{O}[B]$}
\label{table-1}
\begin{tabular}{ | m{7em} | m{12em}|}
\hline \vspace{0.3cm}
$(g\wedge g)_0$ &  Field theory side\\
\hline \vspace{0.3cm}
 $c^{\dagger I}c_{M}\wedge c^{\dagger J}c_{N} $ & $\Psi_{[MN]}^{(IJ)}$ \\
\hline \vspace{0.3cm}
$a^{\dagger}_{\rho}c_{M}\wedge c^{\dagger J} c_{N}$ & $(q_{A\rho})\Psi_{[MN]}^{(IJ)}$ \\
\hline \vspace{0.3cm}
$a^{\dagger}_{\rho}c_{M}\wedge a^\dagger_{\varrho}c_{N}$ & $(q_{B\varrho})(q_{A\rho})\Psi_{[MN]}^{(IJ)}$ \\ 
\hline \vspace{0.3cm}
$0$ & $(q_{[D|\varpi|})(q_{B|\varrho|})(q_{A]\rho})\Psi_{[MN]}^{(IJ)}$\\
\hline \vspace{0.3cm}
$b^\dagger_{\dot\alpha}c^{\dagger I}\wedge \;c^{\dagger J}c_{N} $ & $(\bar q^{A}_{\dot\alpha})\Psi_{[MN]}^{(IJ)}$  \\
\hline \vspace{0.3cm}
$b^\dagger_{\dot\alpha}c^{\dagger I}\wedge \; b^\dagger_{\dot\beta}c^{\dagger J}$ & $(\bar q^{B}_{\dot\beta})(\bar q^{A}_{\dot\alpha})\Psi_{[MN]}^{(IJ)}$ \\ 
\hline \vspace{0.3cm}
$0$ & $(\bar q^{[D}_{\dot\rho})(\bar q^{B}_{\dot\beta})(\bar q^{A]}_{\dot\alpha})\Psi_{[MN]}^{(IJ)}$\\
\hline \vspace{0.3cm}
$b^{\dot\alpha}c_{M}\wedge  c^{\dagger J}c_N$ & $(\bar s^{\dot\alpha}_{A})\Psi_{[MN]}^{(IJ)}$ \\
\hline \vspace{0.3cm}
$b^{\dot\alpha}c_{M}\wedge b^{\dot\beta}c_{N}$ & $(\bar s^{\dot\beta}_{B})(\bar s^{\dot\alpha}_{A})\Psi_{[MN]}^{(IJ)}$ \\ 
\hline \vspace{0.3cm}
$0$ & $(\bar s^{\dot\rho}_{D})(\bar s^{\dot\beta}_{B})( s^{\dot\alpha}_{A})\Psi_{[MN]}^{(IJ)}$\\
\hline \vspace{0.3cm}
$a^{\alpha}c^{\dagger I}\wedge  c^{\dagger J}c_N$ & $( s^{\alpha A})\Psi_{[MN]}^{(IJ)}$ \\
\hline \vspace{0.3cm}
$a^{\alpha}c^{\dagger I}\wedge a^{\beta} c^{\dagger J}$ & $( s^{\beta B})( s^{\alpha A})\Psi_{[MN]}^{(IJ)}$ \\
\hline \vspace{0.3cm}
$0$ & $( s^{\rho D})(s^{\beta B})(s^{\alpha A})\Psi_{[MN]}^{(IJ)}$\\
\hline
\end{tabular}
\end{table}
%
%
On the left hand side we listed  the representations of $psl(4,4|\bf{R})$ namely $(g\wedge g)_0$. 
The AdS/CFT correspondence implies that those representations are related to the field theory by a certain intertwining 
operator.  In the field theory side 
representations of the algebra comes from the subspace in the tensor product of three singletons.

\subsubsection{Action of $q$ and $\bar q$ on $\mathcal{O}[B]$}
Here and the following subsection we describe  the supermultiplet of $\beta-$deformation in the field theory side.

From the supersymmetry algebra, we know that $q_{A\rho}$ and $\bar q^A_{\dot\rho}$ commute with $P_{\mu}$. They act just on $\Psi^{(IJ)}_{[MN]}$.

To begin with, $q_{A\rho}\Psi^{(IJ)}_{[MN]}$ is %
\begin{align}
\label{qpsi}
  q_{A\rho}\Psi^{(IJ)}_{[MN]}\;=\; (c_A a^{\dagger}_{\rho}) \Psi^{(IJ)}_{[MN]} \;\equiv\;&
   2 \delta_A^{(I}\;a^{\dagger}_{\rho} a^{\dagger}_{[1} |0\rangle \bullet a^{\dagger}_{2]} c^{\dagger J)} |0\rangle \bullet \epsilon_{MNKL} c^{\dagger K} c^{\dagger L} |0\rangle \nonumber 
   \\    
   & + 2 \;a^{\dagger}_{[1} c^{\dagger (I}|0\rangle \bullet a^{\dagger}_{2]} c^{\dagger J)} |0\rangle \bullet \epsilon_{AMNL} a^{\dagger}_{\rho} c^{\dagger L} |0\rangle,
\end{align}%

where the right hand side (rhs) is 
\begin{eqnarray} 
  2 \delta_A^{(I}\;f_{\rho[1}\;\psi_{2]}^{J)}\phi_{MN}+2\epsilon_{AMNL}\psi^{(I}_{[1}\psi^{J)}_{2]}\psi^L_{\rho},
 \end{eqnarray}
in terms of oscillators.  To be more clear, what we get is %
\begin{eqnarray}
\label{O-Arho}
 \mathcal{O}_{A\rho}&\equiv& \;\int d^4x B^{[MN]}_{(IJ)}\prod_{j=1}^3 \mathcal{R}^{(j)}q_{A\rho}\Psi^{(IJ)}_{[MN]}  \nonumber \\
 &  =& \int d^4xB^{[MN]}_{(IJ)} ( \delta_A^{I}\;\epsilon^{\alpha\beta}f_{\rho\beta}\;\psi_{\alpha}^{J}\phi_{MN}+\epsilon^{\alpha\beta}\epsilon_{AMNL}\psi^{I}_{\beta}\psi^{J}_{\alpha}\psi^L_{\rho}).
\end{eqnarray}

The result of acting  with a second SUSY generator $q_{B\varrho}$ on ~\eqref{qpsi} is 
\begin{align}
\label{qqpsi}
   (c_B a^{\dagger}_{\varrho})(c_A a^{\dagger}_{\rho}) \Psi^{(IJ)}_{[MN]} \;=\;&
   2 \delta_A^{(I}\;a^{\dagger}_{\rho} a^{\dagger}_{[1} |0\rangle \bullet a^{\dagger}_{2]}  a^{\dagger}_{\varrho}\; \delta^{J)}_{B} |0\rangle \bullet \epsilon_{MNKL} c^{\dagger K} c^{\dagger L} |0\rangle 
    \nonumber \\ &-4 \delta_A^{(I}\;a^{\dagger}_{\rho} a^{\dagger}_{[1} |0\rangle \bullet a^{\dagger}_{2]} c^{\dagger J)} |0\rangle \bullet \epsilon_{BMNL} a^{\dagger}_{\varrho}c^{\dagger L} |0\rangle \nonumber 
   \\    
   & + 4 \delta^{(I}_{B}\;a^{\dagger}_{[1} a^{\dagger}_{\varrho}|0\rangle \bullet a^{\dagger}_{2]} c^{\dagger J)} |0\rangle \bullet \epsilon_{AMNL} a^{\dagger}_{\rho} c^{\dagger L} |0\rangle 
   \nonumber \\ &+ 2 \;a^{\dagger}_{[1} c^{\dagger (I}|0\rangle \bullet a^{\dagger}_{2]} c^{\dagger J)} |0\rangle \bullet \epsilon_{ABMN} a^{\dagger}_{\rho} a^{\dagger}_{\varrho} |0\rangle,
\end{align}%
it leads us to the following
\begin{eqnarray}
\label{O-BvarrhoArho}
\mathcal{O}_{(\varrho B),(\rho A)} &\equiv& \int d^4x  B^{[MN]}_{(IJ)}\prod_{j=1}^3 \mathcal{R}^{(j)}  (c_B a^{\dagger}_{\varrho})(c_A a^{\dagger}_{\rho}) \Psi^{(IJ)}_{[MN]} \nonumber \\
&=&\int d^4x B^{[MN]}_{(IJ)}\epsilon^{\alpha\beta}\big[\delta^{(I}_{A}\delta^{J)}_{B}f_{\rho\beta}f_{\alpha \varrho}\phi_{MN}-2 \epsilon_{BMNL}\delta^{(I}_{A}f_{\rho\beta}\psi_{\alpha}^{J)}\psi_{\varrho}^{L}\nonumber \\ 
&+&2\epsilon_{AMNL}\delta^{(I}_{B}f_{\varrho\beta}\psi^{J)}_{\alpha}\psi^{L}_{\rho}+\epsilon_{ABMN}\psi_{\beta}^{(I}\psi_{\alpha}^{J)}f_{\rho\varrho}\big],
\end{eqnarray}
where the terms in brackets come from~\eqref{qqpsi}.%

\vspace{.3cm}
\underline{Symmetry of $qq\Psi$}%

Antisymmetrization in $A$ and $B$ on~\eqref{qqpsi} gives us%
\begin{align}
 \epsilon^{PQAB}(c_B a^{\dagger}_{\varrho})(c_A a^{\dagger}_{\rho}) \Psi^{(IJ)}_{[MN]}\;=\;0+\text{mod}(\Psi^{(IJ)}_{[MJ]}), \nonumber 
\end{align}
this means that 
\begin{align}
 q_{[A|(\rho|} q_{|B]|\varrho)} \Psi^{(IJ)}_{[MN]}\; =\;  0+\text{traces} .
\end{align}

The above result implies that 
\begin{eqnarray}
 q_{[D|\varpi|}q_{B|\varrho|}q_{A]\rho}\Psi^{(IJ)}_{[MN]}&=&0.
\end{eqnarray}
Therefore we conclude that there are just two descendants~\eqref{O-Arho} and~\eqref{O-BvarrhoArho} obtained from~\eqref{B-multiplet} by applying once and twice $q$, respectively.

\vspace{.3cm}
\underline{Acting with $\bar q$ on $\Psi^{(IJ)}_{[MN]}$}

The action of $\bar q^A_{\dot\alpha}$ and $\bar q^B_{\dot\beta}\bar q^A_{\dot\alpha}$ on $\Psi^{(IJ)}_{[MN]}$ reads as follow
%
\begin{align}
\label{barqpsi}
   (c^{\dagger A} b^{\dagger}_{\dot\alpha}) \Psi^{(IJ)}_{[MN]} \;=\;&
   2 \;b^{\dagger}_{\dot\alpha} a^{\dagger}_{[1}c^{\dagger A}c^{\dagger I} |0\rangle \bullet a^{\dagger}_{2]} c^{\dagger J)} |0\rangle \bullet \epsilon_{MNKL} c^{\dagger K} c^{\dagger L} |0\rangle \nonumber 
   \\    
   & +  \;a^{\dagger}_{[1} c^{\dagger (I}|0\rangle \bullet a^{\dagger}_{2]} c^{\dagger J)} |0\rangle \bullet \epsilon_{MNKL} b^{\dagger}_{\dot\alpha} c^{\dagger A}c^{\dagger K}c^{\dagger L} |0\rangle,
\end{align}
and
\begin{align}
\label{barqbarqpsi}
    (c^{\dagger B} b^{\dagger}_{\dot\beta})(c^{\dagger A} b^{\dagger}_{\dot\alpha}) \Psi^{(IJ)}_{[MN]} \;=\;&
   2 \;b^{\dagger}_{\dot\beta}b^{\dagger}_{\dot\alpha} a^{\dagger}_{[1}c^{\dagger B}c^{\dagger A}c^{\dagger (I} |0\rangle \bullet a^{\dagger}_{2]} c^{\dagger J)} |0\rangle \bullet \epsilon_{MNKL} c^{\dagger K} c^{\dagger L} |0\rangle \nonumber
   \\ 
   %
   &-2 \;b^{\dagger}_{\dot\alpha} a^{\dagger}_{[1}c^{\dagger A}c^{\dagger (I} |0\rangle \bullet a^{\dagger}_{2]} c^{\dagger J)} |0\rangle \bullet \epsilon_{MNKL} b^{\dagger}_{\dot\beta} c^{\dagger B}c^{\dagger K} c^{\dagger L} |0\rangle \nonumber
   \\    
   & +  \;2a^{\dagger}_{[1}b^{\dagger}_{\dot\beta}c^{\dagger B} c^{\dagger (I}|0\rangle \bullet a^{\dagger}_{2]} c^{\dagger J)} |0\rangle \bullet \epsilon_{MNKL} b^{\dagger}_{\dot\alpha} c^{\dagger A}c^{\dagger K}c^{\dagger L} |0\rangle \nonumber 
   \\
   &+a^{\dagger}_{[1} c^{\dagger (I}|0\rangle \bullet a^{\dagger}_{2]} c^{\dagger J)} |0\rangle \bullet \epsilon_{MNKL}b^{\dagger}_{\dot\beta} b^{\dagger}_{\dot\alpha} c^{\dagger B} c^{\dagger A}c^{\dagger K}c^{\dagger L} |0\rangle,
\end{align}
where  we have used the momentum conservation and on-shell condition to get~\eqref{barqbarqpsi}. Those results lead us to 
\begin{eqnarray}
\label{O-dotalpha*A}
 \mathcal{O}^{A}_{\dot\alpha}&\equiv&\int d^4x  \prod^3_{i=1} \mathcal{R}^{(i)} B^{[MN]}_{(IJ)}(c^{\dagger A} b^{\dagger}_{\dot\alpha}) \Psi^{(IJ)}_{[MN]} \nonumber  \\
 &=&\int d^4x B^{[MN]}_{(IJ)}\epsilon^{\alpha\beta} [\partial_{\beta\dot\alpha}\phi^{AI}\psi_{\alpha}^J\phi_{MN}+\frac{1}{2}\psi^{I}_{\beta}\psi^{J}_{\alpha}\delta^{[PA]}_{MN}\bar \psi_{ P \dot\alpha}], 
\end{eqnarray}
and
\begin{eqnarray}
\label{O-dotbetadotalpha*BA}
 \mathcal{O}^{BA}_{\dot\beta\dot\alpha}&\equiv& \int d^4x  \prod^3_{i=1} \mathcal{R}^{(i)} B^{[MN]}_{(IJ)} (c^{\dagger B} b^{\dagger}_{\dot\beta})(c^{\dagger A} b^{\dagger}_{\dot\alpha}) \Psi^{(IJ)}_{[MN]} \nonumber \\
 &=& \int d^4x B^{[MN]}_{(IJ)}\epsilon^{\alpha\beta}\big[\epsilon^{ABPI}\partial_{\beta(\dot\alpha}\bar\psi_{\dot\beta)P}\psi^{J}_{\alpha}\phi_{MN}-\partial_{\beta\dot\alpha|}\phi^{AI}\psi^{J}_{\alpha}\delta^{PB}_{MN}\bar\psi_{\dot\beta P} \nonumber \\
 &+& \partial_{\beta\dot\beta|}\phi^{BI}\psi^{J}_{\alpha}\delta^{PA}_{MN}\bar\psi_{\dot\alpha P}+\frac{1}{2}\psi^{I}_{\beta}\psi^{J}_{\alpha}\delta^{AB}_{MN}\bar f_{(\dot\beta\dot\alpha)}\big],
\end{eqnarray}
where the terms in brackets of~\eqref{O-dotalpha*A} and~\eqref{O-dotbetadotalpha*BA} come from~\eqref{barqpsi} and~\eqref{barqbarqpsi}, respectively.
%
%
%

Acting with three times $\bar q$ on $\Psi$, {\it i.e.} $(c^{\dagger D} b^{\dagger}_{\dot\rho})(c^{\dagger B} b^{\dagger}_{\dot\beta})(c^{\dagger A} b^{\dagger}_{\dot\alpha}) \Psi^{(IJ)}_{[MN]}$, following an antisymmetrization in $A$, $B$ and $D$, we got 
\begin{align}
 (c^{\dagger [D} b^{\dagger}_{(\dot\rho})(c^{\dagger B} b^{\dagger}_{\dot\beta})(c^{\dagger A]} b^{\dagger}_{\dot\alpha)}) \Psi^{(IJ)}_{[MN]} \;=\;0,
\end{align}
this implies that the only descendants obtained from $\mathcal{O}[B]$ after applying once and twice $\bar q $ are~\eqref{O-dotalpha*A} and\eqref{O-dotbetadotalpha*BA}.

These computations were almost straightforward since $q$ and $\bar q$ commute with $\mathcal{R}^{(j)}$. This will not happen with $s^{A\alpha }$ and $\bar s^{\dot\alpha}_{A}$, since the commutator of them with $\mathcal{R}^{(j)}$ is non-zero.
\subsubsection{Action of $s$ and $\bar s$ onto $\mathcal{O}[B]$}
The commutator of $\bar s$ with $p_{\mu}$ is proportional to $q$,  it implies the following result%
%
\begin{align}
\label{[s,R^j]}
[\bar s^{\dot\alpha}_{A}, \mathcal{R}^{(j)}]=-\frac{i}{2}\mathcal{R}^{(j)}x^{\dot\alpha\omega}a^{\dagger}_{\omega}c_{A},
\end{align}
where $x^{\dot\alpha\omega}$ is the contraction $\bar\sigma_{\mu}^{ \dot\alpha\omega}x^{\mu}$. Therefore the action of  $\bar s^{\dot\alpha}_{A}$ on $\mathcal{O}[B]$ is %
\begin{align}
\label{barsO}
 \tilde{\mathcal{O}}^{\dot\alpha}_{A}\;\equiv\;(b^{\dot\alpha}c_{A})\mathcal{O}[B] \;=\;& -\frac{i}{2}\int d^4x \prod_{j=1}^3 \mathcal{R}^{(j)} B^{[MN]}_{(IJ)}x^{\dot\alpha\omega}(a^\dagger_{\omega}c_A)\Psi^{(IJ)}_{[MN]},
\end{align}
where the rhs is 
\begin{align}
 -i\int d^4x B^{[MN]}_{(IJ)}x^{\dot\alpha\omega}\epsilon^{\alpha\beta}\{  \delta_A^{I}\;f_{\omega\beta}\;\psi_{\alpha}^{J}\phi_{MN}+\epsilon_{AMNL}\psi^{I}_{\beta}\psi^{J}_{\alpha}\psi^L_{\omega}\},
\end{align}
in terms of oscillators. 

The result of acting $\bar s$ twice on $\mathcal{O}[B]$ is 
\begin{align}
 \label{barsbarsO}
 \tilde{\mathcal{O}}^{\dot\beta\dot\alpha}_{BA}\;\equiv\;(b^{\dot\beta}c_{B}) (b^{\dot\alpha}c_{A})\mathcal{O}[B] \;=\;&-\frac{1}{4} \int d^4x \prod_{j=1}^3 \mathcal{R}^{(j)} B^{[MN]}_{(IJ)}x^{\dot\alpha\omega}x^{\dot\beta\rho}(a^\dagger_{\rho}c_B)(a^\dagger_{\omega}c_A)\Psi^{(IJ)}_{[MN]},
\end{align}
where the rhs is 
\begin{align}
 -\int d^4x B^{[MN]}_{(IJ)}x^{\dot\alpha\omega}x^{\dot\beta\rho}\epsilon^{\alpha\beta}\{\delta^{I}_{A}\delta^{J}_{B}f_{\omega\beta}f_{\alpha\rho}\phi_{MN}-2 \epsilon_{BMNL}\delta^{I}_{A}f_{\omega\beta}\psi_{\alpha}^{J}\psi_{\rho}^{L} \nonumber \\
 \;+\;2\epsilon_{AMNL}\delta^{I}_{B}f_{\rho\beta}\psi^{J}_{\alpha}\psi^{L}_{\omega}+\epsilon_{ABMN}\psi_{\beta}^{I}\psi_{\alpha}^{J}f_{\omega\rho}\}.
\end{align}

\vspace{.3cm}
\underline{Symmetry of $\bar s \bar s \mathcal{O}[B]$}

Antisymmetrization in $A$ and $B$ on ~\eqref{barsbarsO} lead us to
\begin{align}
\bar s^{(\dot\beta}_{[B}\bar s^{\dot\alpha)}_{A]}\mathcal{O}[B]=0,
\end{align}
it implies that 
\begin{align}
\bar s^{(\dot\varrho}_{[D} \bar s^{\dot\beta}_{B}\bar s^{\dot\alpha)}_{A]}\mathcal{O}[B]=0 .
\end{align}
Therefore, we conclude that the only descendants  obtained from ~\eqref{B-multiplet} after applying once a twice the generator $\bar s $ are ~\eqref{barsO} and ~\eqref{barsbarsO}, respectively.%
 
\vspace{.3cm}
\underline{Acting $s$ on $\mathcal{O}[B]$}

The action of the superconformal generator $s^{A\alpha}$ on  $\mathcal{O}[B]$ reads as follows
%
\begin{eqnarray}
\label{sO}
 \tilde{\mathcal{O}}^{\alpha A}&\equiv& (a^{\alpha}c^{\dagger A})\mathcal{O}[B] \nonumber \\
 &= & \int d^4x \prod^{3}_{j=1}\mathcal{R}^{(j)}B^{[MN]}_{(IJ)}\Big( -ix^{\dot\omega\alpha}a^{\dagger}_{[1} b^\dagger_{\dot\omega}c^{\dagger A} c^{\dagger (I}|0\rangle \bullet a^{\dagger}_{2]} c^{\dagger J)} |0\rangle \bullet \epsilon_{MNKL} c^{\dagger K} c^{\dagger L} |0\rangle \nonumber 
 \\
 &-&\frac{i}{2} a^{\dagger}_{[1} c^{\dagger (I}|0\rangle \bullet a^{\dagger}_{2]} c^{\dagger J)} |0\rangle \bullet \epsilon_{MNKL} x^{\dot\omega\alpha} b^\dagger_{\dot\omega} c^{\dagger A} c^{\dagger K} c^{\dagger L} |0\rangle \nonumber 
 \\
 &+& 2 \delta^{\alpha}_{\;[1}\;c^{\dagger A} c^{\dagger (I}|0\rangle \bullet a^{\dagger}_{2]} c^{\dagger J)} |0\rangle \bullet \epsilon_{MNKL} c^{\dagger K} c^{\dagger L} |0\rangle
 \Big),
\end{eqnarray}
where $x^{\dot\omega\alpha}$ comes from the commutator of $s^{A\alpha}$ with $\mathcal{R}^{(j)}$. The rhs of Eqn. ~\eqref{sO} is just %
\begin{align}
 \int d^4x B^{[MN]}_{(IJ)}\epsilon^{\beta\rho}\{\delta^{\alpha}_{\;\rho}\phi^{AI}\psi^{J}_{\beta}\phi_{MN}-\frac{i}{2}x^{\dot\omega\alpha}(\partial_{\rho\dot\omega}\phi^{AI}\psi_{\beta}^{J}\phi_{MN}+\frac{1}{2}\psi^{I}_{\beta}\psi^{J}_{\rho}\delta_{MN}^{PA}\bar \psi_{P\dot\omega})\},
\end{align}
in the oscillator picture.%

$\tilde{\mathcal{O}}^{\alpha A}$~\eqref{sO} is a descendant obtained from ~\eqref{B-multiplet} and also $s^{B\varrho}s^{A\alpha}\mathcal{O}[B]$, say 
$\tilde{\mathcal{O}}^{(B\varrho),(A\alpha)}$, is. However, the structure of $\tilde{\mathcal{O}}^{(B\varrho),(A\alpha)}$ is not showed explicitly here.They
are just the two descendants from~\eqref{B-multiplet}, since it can be proved that 
\begin{align}
 s^{[D|\varpi|}s^{B|\varrho|}s^{A]\alpha}\mathcal{O}[B]\;=\;0.
\end{align}
Those results described the structure of the supermultiplet of deformation on the field theory side and they were listed in the Table ~\ref{table-1}. 

However we state that still there are another descendant multiplets which can be obtained from ~\eqref{B-multiplet}. Bellow, we list more results .

\begin{small}
\label{llll}
\begin{center}
\begin{tabular}{ | m{12em} | m{12em}|}
\hline \vspace{0.3cm}
$(g\wedge g)_0$ &  Field theory side\\
\hline \vspace{.3cm}
$a^{\dagger}_{\varrho}b^\dagger_{\dot\alpha}\wedge c^{\dagger J}c_N-b^\dagger_{\dot\alpha}c^{\dagger I}\wedge  a^{\dagger}_{\varrho}c_N$ & $(q_{B\varrho})(\bar q^{A}_{\dot\alpha})\Psi_{[MN]}^{(IJ)}$ \\
\hline \vspace{0.3cm}
$\; \; \; \; \; \;b^\dagger_{\dot\alpha}a^{\dagger}_{[\varrho}\wedge  a^\dagger_{\varpi]}c_N$ &$(q_{C\varpi})(q_{B\varrho})(\bar q^{A}_{\dot\alpha})\Psi_{[MN]}^{(IJ)}$\\
\hline \vspace{0.4cm}
$\;\;\;\;\;\;\;0$ & $(q_{D\vartheta})(q_{C\varpi})(q_{B\varrho})(\bar q^{A}_{\dot\alpha})\Psi_{[MN]}^{(IJ)}$ \\
\hline \vspace{0.3cm}
$a^{\dagger}_{\varrho}b^\dagger_{\dot\alpha}\wedge b^\dagger_{\dot\beta}c^{\dagger J}-b^{\dagger}_{\dot\alpha}c^{\dagger I}\wedge a^\dagger_{\varrho} b^\dagger_{\dot\beta}$ & $(q_{D\varrho})(\bar q^{B}_{\dot\beta})(\bar q^{A}_{\dot\alpha})\Psi_{[MN]}^{(IJ)}$ \\
\hline \vspace{0.3cm}
$\;\;\;\;\;\;b^\dagger_{\dot\alpha}a^{\dagger}_{[\varrho}\wedge a^\dagger_{\varpi]} b^\dagger_{\dot\beta}$ & $(q_{E\varpi})(q_{D\varrho})(\bar q^{B}_{\dot\beta})(\bar q^{A}_{\dot\alpha})\Psi_{[MN]}^{(IJ)}$\\
\hline \vspace{0.3cm}
$\;\;\;\;\;\;0$ & $q\;q\;q (\bar q^{B}_{\dot\beta})(\bar q^{A}_{\dot\alpha})\Psi_{[MN]}^{(IJ)}$\\
 \hline \vspace{0.3cm}
 $a^{\alpha}b^{\dot\alpha}\wedge c_N c^{\dagger J}-b^{\dot\alpha}c_M\wedge a^{\alpha} c^{\dagger J}$& $(s^{\alpha B})(\bar s^{\dot\alpha}_{A})\Psi^{(IJ)}_{[MN]}$ \\
 \hline \vspace{0.3cm}
 $\;\;\;\;\;\;b^{\dot\alpha}a^{[\alpha}\wedge a^{\beta]} c^{\dagger J}$& $(s^{\beta D})(s^{\alpha B})(\bar s^{\dot\alpha}_{A})\Psi^{(IJ)}_{[MN]}$ \\
 \hline \vspace{0.3cm}
 $\;\;\;\;\;\;\;0$ & $(s^{\rho E})(s^{\beta D})(s^{\alpha B})(\bar s^{\dot\alpha}_{A})\Psi^{(IJ)}_{[MN]}$ \\
 \hline \vspace{0.3cm}
 $a^{\alpha}b^{\dot\alpha}\wedge b^{\dot\beta}c_N-b^{\dot\alpha}c_M\wedge a^{\alpha}b^{\dot\beta}$ & $(s^{\alpha D})(\bar s^{\dot\beta}_{B})(\bar s^{\dot\alpha}_{A})\Psi^{(IJ)}_{[MN]}$ \\
 \hline \vspace{0.3cm}
 $\;\;\;\;\;\;b^{\dot\alpha}a^{[\alpha}\wedge a^{\beta]} b^{\dot\beta}$ & $(s^{\beta E})(s^{\alpha D})(\bar s^{\dot\beta}_{B})(\bar s^{\dot\alpha}_{A})\Psi^{(IJ)}_{[MN]}$\\
 \hline  \vspace{0.3cm}
 $\;\;\;\;\;\;0$& $sss(\bar s^{\dot\beta}_{B})(\bar s^{\dot\alpha}_{A})\Psi^{(IJ)}_{[MN]}$\\
\hline \vspace{0.3cm}
 $ b_{\dot\beta}^{\dagger} b^{\dot\alpha}\wedge  c^{\dagger J}c_{N}-b^{\dot\alpha}c_M\wedge b^\dagger_{\dot\beta}c^{\dagger J}$ & $(\bar q^{B}_{\dot\beta})(\bar s^{\dot\alpha}_{A})\Psi^{(IJ)}_{[MN]}$ \\
 \hline \vspace{0.3cm}
$\;\;\;\;\;\;b_{[\dot\beta}^{\dagger} b^{\dot\alpha}\wedge b^\dagger_{\dot\rho]}c^{\dagger J}$ & $(\bar q^{D}_{\dot\rho})(\bar q^{B}_{\dot\beta})(\bar s^{\dot\alpha}_{A})\Psi^{(IJ)}_{[MN]}$\\
 \hline \vspace{0.3cm}
 $\;\;\;\;\;\;0$ & $(\bar q^{E}_{\dot\varrho})(\bar q^{D}_{\dot\rho})(\bar q^{B}_{\dot\beta})(\bar s^{\dot\alpha}_{A})\Psi^{(IJ)}_{[MN]}$\\
 \hline \vspace{0.3cm}
 $b^\dagger_{\dot\rho}b^{\dot\alpha}\wedge b^{\dot\beta}c_N-b^{\dot\alpha}c_M\wedge b^\dagger_{\dot\rho}b^{\dot\beta}$& $(\bar q^{D}_{\dot\rho})(\bar s^{\dot\beta}_{B})(\bar s^{\dot\alpha}_{A})\Psi_{[MN]}^{(IJ)}$\\
 \hline \vspace{0.3cm}
 $\;\;\;\;\;\;b^\dagger_{[\dot\rho}b^{\dot\alpha}\wedge b^\dagger_{\dot\varrho]}b^{\dot\beta}$& $(\bar q^{E}_{\dot\varrho})(\bar q^{D}_{\dot\rho})(\bar s^{\dot\beta}_{B})(\bar s^{\dot\alpha}_{A})\Psi_{[MN]}^{(IJ)}$\\
 \hline \vspace{0.3cm}\
 $\;\;\;\;\;\;0$ & $\bar q \bar q \bar q (\bar s^{\dot\beta}_{B})(\bar s^{\dot\alpha}_{A})\Psi_{[MN]}^{(IJ)}$\\
 \hline \vspace{0.3cm}
 $a^\dagger_{\beta}a^{\alpha}\wedge c^{\dagger J}c_N -a^{\alpha}c^{\dagger I}\wedge a^\dagger_{\beta}c_N $ & $(q_{B\beta})( s^{\alpha A})\Psi^{(IJ)}_{[MN]}$\\
 \hline \vspace{0.3cm}
 $\;\;\;\;\;\;a^\dagger_{[\beta}a^{\alpha}\wedge a^\dagger_{\rho]}c_N$ & $(q_{D\rho})(q_{B\beta})( s^{\alpha A})\Psi^{(IJ)}_{[MN]}$\\
 \hline \vspace{0.3cm}
 $\;\;\;\;\;\;0$ & $(q_{E\varrho})(q_{D\rho})(q_{B\beta})( s^{\alpha A})\Psi^{(IJ)}_{[MN]}$\\
 \hline \vspace{0.3cm}
 $a^\dagger_{\rho}a^{\alpha}\wedge a^{\beta} c^{\dagger J}-a^{\alpha}c^{\dagger I}\wedge a^\dagger_{\rho}a^{\beta} $ & $(q_{D\rho})( s^{\beta B})( s^{\alpha A})\Psi^{(IJ)}_{[MN]}$\\
 \hline \vspace{0.3cm}
 $\;\;\;\;\;\;a^\dagger_{[\rho}a^{\alpha}\wedge a^{\dagger}_{\varrho]}a^{\beta}$ & $(q_{E\varrho})(q_{D\rho})( s^{\beta B})( s^{\alpha A})\Psi^{(IJ)}_{[MN]}$\\
 \hline \vspace{0.3cm}
 $\;\;\;\;\;\;0$ & $qqq( s^{\beta B})( s^{\alpha A})\Psi^{(IJ)}_{[MN]}$\\
 \hline
\end{tabular}
\end{center}
\end{small}

The results listed in this table are possible, too. For instance, let us choose 
\begin{align}
 sss(\bar s^{\dot\beta}_B)(\bar s^{\dot\alpha}_A)\Psi^{(IJ)}_{[MN]},
\end{align}
using relation of commutations. It can be written as 
\begin{align}
\label{sssbarsbarspsi-discussion}
 ss(\bar s^{\dot\beta}_B)(\bar s^{\dot\alpha}_A)s^{\star\circ}\Psi^{(IJ)}_{[MN]}-\delta^{\star}_A ss(\bar s^{\dot\beta}_B) k^{\dot\alpha\circ}\Psi^{(IJ)}_{[MN]}+\delta^{\star}_Bssk^{\dot\beta\circ}(\bar s^{\dot\alpha}_A)\Psi^{(IJ)}_{[MN]} ,
 \end{align}
let us recall that  the complete expresion in the field theory side is~\eqref{B-multiplet}, for a brievity let us say $\Psi^{(IJ)}_{[MN]}$.
In Appendix \ref{a} we check that~\eqref{B-multiplet} is invariant under the generator of conformal transformations $k^{\dot\alpha\alpha }$. Moreover the generators $k$ and $\bar s$ commutes. Therefore the last two terms in~\eqref{sssbarsbarspsi-discussion}
vanish. With this procedure we have the following equality
\begin{align}
  sss(\bar s^{\dot\beta}_B)(\bar s^{\dot\alpha}_A)\Psi^{(IJ)}_{[MN]}\;=\;ss(\bar s^{\dot\beta}_B)(\bar s^{\dot\alpha}_A)s^{\star\circ}\Psi^{(IJ)}_{[MN]}\;=\; (\bar s^{\dot\beta}_B)(\bar s^{\dot\alpha}_A)[sss\Psi^{(IJ)}_{[MN]}]\;=\;0,
\end{align}
where we have used the result listed in table \ref{table-1}.

With the same reasoning we get additional representations of the $psl(4|4,\bf{R})$ algebra, in  subspaces of the tensor product of 
three singleton representations.%

\vspace{.7cm}
{\bf{Acknowledgments}}

I am gratefull to Prof. Andrei Mikhailov for suggesting the problem and for his advice. The current work 
was supported by  the
Brazilian CNPq scholarship and in part by FAPESP grant 2014/18634-9.  
%
%
%
%
%
%
\appendix
\section{$\mathcal{O}[B]$ is conformal invariant}
\label{a}
Here we check that $\mathcal{O}[B]$ is invariant under conformal transformations. To do that 
we act by the generator of conformal transformation $k^{\dot\varrho\varrho}$ on it, first we need the following result 
\begin{align}
\label{kR}
 [a^{\varrho}b^{\dot\varrho}, \mathcal{R}^{(j)}]\;=\; -\frac{i}{2}\mathcal{R}^{(j)}\big(x^{\varrho\alpha'}a^\dagger_{\alpha'}a^{\varrho}+ x^{\dot\alpha\varrho}b^\dagger_{\dot\alpha}b^{\dot\varrho}+x^{\dot\varrho\varrho}-\frac{i}{2}x^{\dot\varrho\alpha'}x^{\dot\rho\varrho}a^\dagger_{\alpha'}b^\dagger_{\dot\rho}\big).
\end{align}
Therefore $(a^{\varrho}b^{\dot\varrho})\mathcal{O}[B]$ is given by

\begin{align}
 &\int d^4x \Bigg( \Big(\big(\frac{3}{2}\frac{\partial}{\partial p^{(1)}_{\varrho\dot\varrho}}\prod_{i=1}^3 \mathcal{R}^{(i)} +\{\frac{\partial}{\partial p^{(1)}_{\alpha'\dot\varrho}}\frac{\partial}{\partial p^{(1)}_{\varrho\dot\rho}}\prod_{i=1}^3 \mathcal{R}^{(i)}\}p^{(1)}_{\alpha'\dot\rho}\big) c^{A\dagger}a_{[1}^{\dagger}|0\rangle  \Big)\bullet c^{B\dagger}a_{2]}^{\dagger}|0\rangle \bullet \epsilon_{ABMN}c^{M\dagger}c^{N\dagger} |0\rangle \nonumber \\ 
 & \;+\;c^{A\dagger}a_{[1}^{\dagger}|0\rangle\bullet\Big(\big(\frac{3}{2}\frac{\partial}{\partial p^{(2)}_{\varrho\dot\varrho}} \prod_{i=1}^3 \mathcal{R}^{(i)} +\{\frac{\partial}{\partial p^{(2)}_{\alpha'\dot\varrho}}\frac{\partial}{\partial p^{(2)}_{\varrho\dot\rho}} \prod_{i=1}^3 \mathcal{R}^{(i)}\}p^{(2)}_{\alpha'\dot\rho}\big) c^{B\dagger}a_{2]}^{\dagger}|0\rangle  \Big) \bullet \epsilon_{ABMN}c^{M\dagger}c^{N\dagger} |0\rangle \nonumber \\ 
 & \;+\;c^{A\dagger}a_{[1}^{\dagger}|0\rangle\bullet c^{B\dagger}a_{2]}^{\dagger}|0\rangle \bullet \big(\frac{\partial}{\partial p^{(3)}_{\varrho\dot\varrho}}\prod_{i=1}^3 \mathcal{R}^{(i)}+\{\frac{\partial}{\partial p^{(3)}_{\alpha'\dot\varrho}}\frac{\partial}{\partial p^{(3)}_{\varrho\dot\rho}}\prod_{i=1}^3 \mathcal{R}^{(i)}\}p^{(3)}_{\alpha'\dot\rho}\big) \epsilon_{ABMN}c^{M\dagger}c^{N\dagger} |0\rangle \Bigg),  \nonumber 
\end{align}
where we have used~\eqref{kR} and antisymmetrization properties. Taking the integral we have %
\begin{align}
\label{trilinear}
   &\Big(\big(\frac{3}{2}\frac{\partial}{\partial p^{(1)}_{\varrho\dot\varrho}}\delta(\sum^3_{i=1}p^{(i)}) +\{\frac{\partial}{\partial p^{(1)}_{\alpha'\dot\varrho}}\frac{\partial}{\partial p^{(1)}_{\varrho\dot\rho}}\delta(\sum_{i=1}^3 p^{(i)})\}p^{(1)}_{\alpha'\dot\rho}\big) c^{A\dagger}a_{[1}^{\dagger}|0\rangle  \Big)\bullet c^{B\dagger}a_{2]}^{\dagger}|0\rangle \bullet \epsilon_{ABMN}c^{M\dagger}c^{N\dagger} |0\rangle \nonumber \\
 & \;+\;c^{A\dagger}a_{[1}^{\dagger}|0\rangle\bullet\Big(\big(\frac{3}{2}\frac{\partial}{\partial p^{(2)}_{\varrho\dot\varrho}} \delta(\sum_{i=1}^3 p^{(i)}) +\{\frac{\partial}{\partial p^{(2)}_{\alpha'\dot\varrho}}\frac{\partial}{\partial p^{(2)}_{\varrho\dot\rho}}\delta(\sum_{i=1}^3 p^{(i)})\}p^{(2)}_{\alpha'\dot\rho}\big) c^{B\dagger}a_{2]}^{\dagger}|0\rangle  \Big) \bullet \epsilon_{ABMN}c^{M\dagger}c^{N\dagger} |0\rangle \nonumber \\
 & \;+\; c^{A\dagger}a_{[1}^{\dagger}|0\rangle\bullet c^{B\dagger}a_{2]}^{\dagger}|0\rangle \bullet \big(\frac{\partial}{\partial p^{(3)}_{\varrho\dot\varrho}}\delta(\sum_{i=1}^3 p^{(i)})+\{\frac{\partial}{\partial p^{(3)}_{\alpha'\dot\varrho}}\frac{\partial}{\partial p^{(3)}_{\varrho\dot\rho}}\delta(\sum_{i=1}^3 p^{(i)})\}p^{(3)}_{\alpha'\dot\rho}\big) \epsilon_{ABMN}c^{M\dagger}c^{N\dagger} |0\rangle.  
\end{align}
From $\delta(\sum_{i=1}^3 p^{(i)})\}\sum_{j=1}^3 p^{(j)}_{\alpha'\dot\rho}=0$, we get the following identity %
\begin{align}
&\frac{1}{4}\sum_{l=1}^3\sum_{j=1}^3g_{lj}\{\frac{\partial}{\partial p^{(l)}_{\alpha'\dot\varrho}}  \frac{\partial}{\partial p^{(j)}_{\varrho \dot\rho}}\delta(\sum_{i=1}^3 p^{(i)})\}\sum_{k=1}^3 p^{(k)}_{\alpha'\dot\rho}+\sum_{k=1}^3\sum_{l=1}^3g_{kl}\{\frac{\partial}{\partial p^{(l)}_{\varrho \dot\varrho}}\delta(\sum_{i=1}^3 p^{(i)})\} \;=\;0, \nonumber 
\end{align}
let  $g_{lm}$ be $a_l\delta^{l}_{m}$, so the last identity reduces to
\begin{align}
\label{trick}
  &\frac{1}{4}\big(\sum_{l=1}^3a_l\big)\{\frac{\partial}{\partial p^{(m)}_{\alpha'\dot\varrho}}  \frac{\partial}{\partial p^{(m)}_{\varrho \dot\rho}}\delta(\sum_{i=1}^3 p^{(i)})\}\sum_{k=1}^3 p^{(k)}_{\alpha'\dot\rho}+\sum_{k=1}^3a_{k}\{\frac{\partial}{\partial p^{(k)}_{\varrho \dot\varrho}}\delta(\sum_{i=1}^3 p^{(i)})\}\;=\;0 ,
\end{align}
there is no sumation on $m$ index, Eqn. ~\eqref{trilinear} can be written as 
\begin{align}
 &\mathcal{O} (c^{A\dagger}a_{[1}^{\dagger}|0\rangle\bullet c^{B\dagger}a_{2]}^{\dagger}|0\rangle \bullet  \epsilon_{ABMN}c^{M\dagger}c^{N\dagger} |0\rangle )
\end{align}
where $\mathcal{O}$ is the identity in  ~\eqref{trick}, with $a_1=a_2=\frac{3}{2}$ and $a_3=1$. Therefore  we have checked that  $\mathcal{O}[B]$ is invariant under conformal transformations.

\end{document}